\documentclass[twoside]{article}
\usepackage[accepted]{aistats2018}
\usepackage{dsfont}
\usepackage{natbib}
\usepackage{amsmath}
\usepackage{amssymb}
\usepackage{enumerate}
\usepackage{graphicx}
\usepackage{subcaption}
\usepackage{tikz}
\usepackage{breqn}
\usetikzlibrary{shapes.geometric}

\usepackage{algpseudocode}
\usepackage{algorithm}

\usepackage{booktabs,colortbl}
\definecolor{Gray}{gray}{0.9}

\newcommand{\ind}[1]{\mathds{1}(#1)}%

\algnewcommand\algorithmicinput{\textbf{Input:}}
\algnewcommand\INPUT{\item[\algorithmicinput]}

\algnewcommand\algorithmicoutput{\textbf{Output:}}
\algnewcommand\OUTPUT{\item[\algorithmicoutput]}

\newcommand{\subjectto}{\mbox{subject to}}                       
\begin{document}

\twocolumn[

\aistatstitle{Practical Bayesian Inference for Record Linkage}

\aistatsauthor{Brendan S. McVeigh \And Jared S. Murray}
\aistatsaddress{Carnegie Mellon University \And University of Texas at Austin}]

\begin{abstract}
Probabilistic record linkage (PRL) is the process of determining which records in two databases correspond to the same underlying entity in the absence of a unique identifier. Bayesian solutions to this problem provide a powerful mechanism for propagating uncertainty due to uncertain links between records (via the posterior distribution).  However, computational considerations severely limit the practical applicability of existing Bayesian approaches. 

We propose a new computational approach, providing both a fast algorithm for deriving point estimates of the linkage structure that properly account for one-to-one matching and a restricted MCMC algorithm that samples from an approximate posterior distribution. Our advances make it possible to perform Bayesian PRL for larger problems, and to assess the sensitivity of results to varying prior specifications. We demonstrate the methods on simulated data and an application to a post-enumeration survey for coverage estimation in the Italian census.
\end{abstract}

\section{INTRODUCTION}
Probabilistic record linkage (PRL) is the task of merging two or more databases that have entities in common but no unique identifier.  In this setting matching records must be done based on incomplete information; features for records may be incorrectly or inconsistently recorded and others may be missing altogether.  Uncertainty in the true matching structure makes this fundamentally a statistical problem. Bayesian approaches to PRL are appealing as they provide a natural mechanism for accounting for this uncertainty via the posterior distribution.
\vspace{0.5em}

Early applications of PRL include linking files from surveys and censuses  to estimate the number of records in common and estimate the total population size in capture-recapture studies \citep{neter1965effect,fellegi1969theory,winkler1991application}. Recent Bayesian approaches to this problem include \citep{liseo2011bayesian, tancredi2011hierarchical, tancredi2013accounting}, among others.  Similar methods have been applied for estimating casualty counts in conflict regions \citep{steorts2016bayesian, ventura2014hierarchical, sadinle2014detecting, sadinle2016bayesian}.
\vspace{0.5em}

PRL is also used to merge files and gather more complete information about individuals, e.g. if covariates and a response are recorded on different files \citep{gutman2013bayesian, gu2016combining, dalzell2016regression}.  Areas of application include linking healthcare data across providers
\citep{dusetzina2014linking, sauleau2005medical} and following students across schools \citep{mackay2015educational, alicandro2017differences}.
\vspace{0.5em}

Here we consider merging two files with no duplicate records, so that each record in the first file matches at most one record in the second (``one-to-one'' matching).  Duplicates and multiple files could be handled using similar methods to ours, but we leave these for future work.
\vspace{0.5em}


\section{A RECORD LINKAGE MODEL}

Suppose we have two collections of records, denoted $A$ and $B$, containing $n_A$ and $n_B$ records (respectively). We assume without loss of generality that $n_A \leq n_B$.  Records $a\in A$ and $b\in B$ are said to be ``matched'' or ``linked'' if they refer to the same underlying entity, which we denote $a\sim b$.  In PRL the object of inference is the set of true links, which can be conveniently represented in matrix form:
\begin{equation}
	C_{ab} = \left\{
\begin{array}{ll}
      1 & a \sim b \\
      0 & a \nsim b. \\
\end{array} 
\right.
\end{equation}

The matrix $C$ is unobserved, but for each record we obtain a set of features (names, addresses, demographic information, and so on) that serve to weakly identify the individual to whom the record belongs.  We assume that our data are in the form of {\em comparisons} of record pairs defined using these features, such as the absolute differences in the age field of two records, or the edit distance between two names (see \cite{christen2012data,herzog2007data} for a detailed account of generating comparisons). Specifically, for the record pair $(a,b)$ we observe a length $d$ vector of comparisons:
\begin{equation}
    \gamma_{ab} = (\gamma_{ab}^1, \gamma_{ab}^2, \dots, \gamma_{ab}^d).
\end{equation}

Each $\gamma_{ab}^j$ is an ordinal level of agreement. For example, when comparing two strings in Section \ref{sec:simstudy} a string similarity score is partitioned into high, medium, moderate, and low levels of agreement, coded as 4, 3, 2, and 1 respectively (as in \cite{sadinle2014detecting}, for example).  The reduction to comparisons reduces the modeling burden significantly, especially in the presence of complex features like strings. Alternative modeling approaches are discussed in Section 6. 

\noindent	
These comparison vectors are modeled using a two component mixture, with realized comparison vectors corresponding to either matching or non-matching record pairs. Specifically, our parameters are given by
\begin{equation}
\begin{split}
m(g) &= \Pr\left(\gamma_{ab} = g\mid C_{ab} = 1\right) \\
u(g) &= \Pr\left(\gamma_{ab} = g\mid C_{ab} = 0\right),
\end{split}
\end{equation}
for $g$ ranging over all possible comparison vectors. 

To reduce the number of parameters in the model we make the common assumption of conditional independence (given $C$) between comparisons. Define
\begin{align}
\begin{split}
m_{jh} &= \Pr\left(\gamma_{ab}^j = h|C_{ab} = 1\right)\\ 
u_{jh} &= \Pr\left(\gamma_{ab}^j = h|C_{ab} = 0\right),
\end{split}
\end{align}
for $1\leq j\leq d$ and $1\leq h\leq k_j$, where comparison $j$ has $k_j$ possible levels.
Under conditional independence:
\begin{equation}
\begin{split}
		m(\gamma_{ab}) = \prod_{j=1}^d \prod_{h=1}^{k_i} m_{jh}^{\ind{\gamma_{ab}^j=h}}\\
		u(\gamma_{ab}) =  \prod_{j=1}^d \prod_{h=1}^{k_i} u_{jh}^{\ind{\gamma_{ab}^j=h}}. 
\end{split}
\end{equation}
Finally, we define a weight for each record pair:
\begin{equation}  \label{eq:weight}
w_{ab} = \log\left(\frac{m(\gamma_{ab})}{u(\gamma_{ab})}\right).
\end{equation}
The weight summarizes information about the \textit{relative} likelihood of a record pair being a link versus non-link.  Informally we can think of $w_{ab}$ as a log-likelihood ratio statistic for testing whether $\gamma_{ab}$ was generated by comparing matching or non-matching records.

\section{POINT ESTIMATES FOR C}

There two common approaches to generating a point estimate $\hat C$ of $C$:  A three-stage procedure introduced by \cite{jaro1989advances}, and fully Bayesian approaches that obtain $\hat C$ by minimizing the expected value of a loss function with respect to a posterior distribution over $C$. It is well known that Bayes estimates are generally superior to the three-stage estimate \citep[see e.g.][]{tancredi2011hierarchical, sadinle2016bayesian}, but incur a greater computational cost. After reviewing these approaches, we introduce a penalized maximum likelihood approach that partially bridges this gap in Section~\ref{sec:penlike}.

\subsection{The Fellegi-Sunter Decision Rule with Constraints}\label{sec:EMLSAP}
\cite{fellegi1969theory} provided an early framework for estimating $C$.  Given weights $w_{ab}$ from \eqref{eq:weight}, $\hat C_{ab}$ is set to 1 if $w_{ab}>\lambda$ and 0 if $w_{ab} < \mu$ for thresholds $\lambda$ and $\mu$ chosen to control the false negative and false positive rate. Any record pairs with $\mu \leq w_{ab}\leq \lambda$ are assigned an indeterminate match status, and possibly sent for additional clerical review (similar to later classification rules that allow a ``reject'' option).

\cite{fellegi1969theory} developed the theory for their decision framework under known $m$ and $u$ values. 
In practice these must be estimated, usually via EM \citep{winkler1988using} under the likelihood
\begin{equation}
L(m,u, \pi; \Gamma) = \prod_{(a,b)\in A\times B}
\pi m(\gamma_{ab}) + (1-\pi) u(\gamma_{ab})\label{eq:maxlik},
\end{equation}
which treats the comparison vectors as independent ``observations'' from a two component mixture model. The estimate $\hat C$ is then derived using the Fellegi-Sunter decision rule, plugging in estimates $\hat m$ and $\hat u$.

Treating the record pairs as independent observations during estimation and in the decision rule can lead to poor or even nonsensical results, particularly under one-to-one matching. \cite{jaro1989advances} developed a three-stage approach for estimating $C$ under one-to-one matching. The first stage generates estimates of $\hat m$ and $\hat u$ by maximizing \eqref{eq:maxlik}. The second stage generates an estimate of $C$ that satisfies the following {\em linear sum assignment problem (LSAP)}:

\begin{equation} \label{eq:jaroLSAP}
\begin{aligned}
\max_{C}   &\sum_{a,b\in A\times B} C_{ab}\hat w_{ab}& \\
\subjectto\quad  & C_{ab} \in\{0, 1\}\\
&\sum_{b\in B} \hspace{0.5em} C_{ab} = 1 \hspace{2em} \forall a\in A \\
&\sum_{a\in A} \hspace{0.5em} C_{ab} = 1 \hspace{2em} \forall b\in B,
\end{aligned}
\end{equation}
In the final stage, \cite{jaro1989advances} obtains $\hat C$ from the solution to the optimization problem $\tilde C$ by setting $\hat C_{ab} = \tilde C_{ab}\ind{\hat w_{ab}>\mu}$, where $\mu$ plays a similar role here as in the Fellegi-Sunter decision rule. We will refer to this as the EM  + LSAP approach. This method continues to be deployed in applications; see e.g. \cite{enamorado2017using} for a recent example.



\subsection{Bayes Estimates}

A Bayesian approach to the PRL problem begins with a prior $p(C)$, which provides a vehicle for imposing constraints like one-to-one matching. In the context of our comparison-based model, the joint posterior distribution for $C$ and the other parameters is
\begin{equation}
p(C, m, u\mid \Gamma) \propto p(C)p(m, u)p(\Gamma\mid C, m, u)
\end{equation}
(assuming $C$ is independent from the other parameters {\em a priori}, as is common). 

Bayes estimates for $C$ can be derived by minimizing the expected loss with respect to the posterior. \cite{tancredi2011hierarchical} show that under squared error or balanced misclassification loss functions the Bayes estimate $\hat C$ is obtained by setting 
\begin{equation}
    \hat C_{ab}=\ind{\Pr(C_{ab}=1\mid \Gamma)> 0.5}.\label{eq:bayesest}
\end{equation}
\cite{sadinle2016bayesian} introduced more sophisticated loss functions that include a ``reject'' or ``indeterminate'' option, similar to the Fellegi-Sunter decision rule in the previous subsection.  

Compared to the EM+LSAP approach, computing a Bayes estimate is expensive as it entails a long MCMC run to get accurate estimates of $\Pr(C_{ab}=1\mid \Gamma)$. However, the EM+LSAP approach can be notoriously inaccurate \citep{tancredi2011hierarchical, sadinle2016bayesian}. In the next section we develop a new penalized likelihood approach that is both computationally efficient and more accurate than EM+LSAP.

\subsection{A New Penalized Likelihood Estimate}\label{sec:penlike}

The ``likelihood'' for $m$ and $u$ in \eqref{eq:maxlik} is equivalent to the marginal likelihood under a model for $C$ that has $\Pr(C_{ab}=1) = \pi$ independently for all $(a,b)\in A\times B$. This implicit model is grossly misspecified under one-to-one matching. 

We could alternatively estimate $C$ by maximizing a joint likelihood in $C$, $m$, and $u$ with the appropriate restrictions. By penalizing this joint likelihood we can also avoid the post-hoc thresholding by $\mu$ in the EM+LSAP procedure. Ignoring regularization of $m$ and $u$, a natural choice for the (log) penalized likelihood is

\begin{dmath}
    \sum_{ab}  [C_{ab}\log(m(\gamma_{ab}) + (1 - C_{ab})\log(u(\gamma_{ab}))]
    -\theta\sum_{ab} C_{ab}\hfill.\label{eq:penlikend}
\end{dmath}
Here $\theta$ is the penalty parameter representing the cost of each additional link. The penalty term is better understood by rearranging \eqref{eq:penlikend} as 
\begin{align}
    l(C, m, u; \Gamma) &= \sum_{ab} \log(u_{ab}) + C_{ab}(w_{ab} -\theta),\label{eq:penlik}
\end{align}
where again $w_{ab} = \log[m(\gamma_{ab})/u(\gamma_{ab})]$. Hence $\theta$ plays a similar role to $\mu$ in the Fellegi-Sunter decision rule; only pairs with $w_{ab}>\theta$ can be linked without decreasing the log-likelihood. However, in our penalized likelihood approach this constraint is enforced during estimation. We will see in Section~\ref{sec:experiments} that this has important implications for parameter estimation.

A local mode of \eqref{eq:penlik} is readily obtained by alternating maximization steps: Holding $m,u$ constant, maximizing \eqref{eq:penlik} in $C$ is equivalent to solving the following optimization problem:

\begin{equation} \label{eq:penlike}
\begin{aligned}
\max_{C}   &\sum_{a,b\in A\times B} C_{ab} (w_{ab} - \theta) \\
\subjectto\quad  & C_{ab} \in\{0, 1\}\\
&\sum_{b\in B} \hspace{0.5em} C_{ab} \leq 1 \hspace{2em} \forall a\in A \\
&\sum_{a\in A} \hspace{0.5em} C_{ab} \leq 1 \hspace{2em} \forall b\in B.
\end{aligned}
\end{equation}
The solution can be found by solving an LSAP and then simply deleting links where $w_{ab} \leq \theta$, details on the LSAP are given in Appendix~\ref{sec:LSAP}.  This is similar to \cite{jaro1989advances}'s approach, except it is one step of an iterative maximization routine here.  Efficient algorithms exist for solving LSAPs, (e.g. the Hungarian algorithm \citep{kuhn1955hungarian}) and have a worst case complexity of $O(n^3)$ where $n = \max(n_A, n_B)$ \citep{jonker1986improving, lawler1976combinatorial}.  (\cite{green2015mad} proposes a similar penalized likelihood approach for alignment problems under a different class of models.)

Holding $C$ constant, \eqref{eq:penlik} separates into distinct factors for each $m_{jh}$ and $u_{jh}$, which are maximized by setting
\begin{align}
m_{jh} &= 
\frac{
n_{mjh} + \sum_{ab} C_{ab}\ind{\gamma^j_{ab} = h}
}{
\sum_h n_{mjh} + \sum_{ab} C_{ab}
}\\
u_{jh} &= 
\frac{
n_{ujh} + \sum_{ab} (1-C_{ab})\ind{\gamma^j_{ab} = h}
}{
\sum_h n_{ujh} + \sum_{ab} (1-C_{ab})
}.
\end{align}
where the $n$'s are optional pseudocounts used to regularize the estimates (omitted from \eqref{eq:penlik}).

\subsection{Comparing Point Estimators}\label{sec:comp_point_est}

The penalized likelihood and EM+LSAP estimators have similar computational complexity; although the penalized likelihood estimator requires solving the LSAP a handful of times, the previous value of the weights can be used as warm-starts to reach a solution quickly. Both are much faster than computing a Bayes estimate using MCMC algorithms.

In terms of accuracy, we will see in Section~\ref{sec:italy} that penalized likelihood estimates can produce point estimates of $C$ that are very similar to Bayes estimates, while EM+LSAP simultaneous produces a questionable point estimate. Further, we will see that the estimates of $m$ and $u$ produced by EM can be badly biased, because of the misspecified marginal likelihood in \eqref{eq:maxlik}.

However, the Bayes estimates have an advantage over both penalized likelihood and EM+LSAP in their treatment of uncertainty. To take an extreme example, if a record $a$ has multiple exact matches $b,\ b'$ in file $B$ then the penalized likelihood and EM+LSAP methods will assign $a$ to either $b$ or $b'$ more or less at random. However, the posterior probabilities $\Pr(C_{ab}=1\mid \Gamma)$ and $\Pr(C_{ab'}=1\mid \Gamma)$ will be approximately equal and less than $0.5$ because the posterior incorporates the one-to-one constraint. Hence record $a$ will generally be either left unmatched, or the pairs $(a,b)$ and $(a,b')$ will be assigned an indeterminate status using e.g. \cite{sadinle2016bayesian}'s loss functions. This behavior seems more desirable in a point estimate. 

More generally, the full posterior distribution provides a natural mechanism for propagating uncertainty in the true match/non-match status through to inference using the linked files. However, methods for approximate Bayesian inference are extremely computationally challenging: There are $n_An_B$ possible links to explore, and in principle a well-mixing MCMC algorithm should explore them all. In the next section we explore how fast, high-quality penalized likelihood estimates can be used to scale approximate posterior inference.

\section{POST-HOC BLOCKING AND RESTRICTED MCMC}

\begin{figure*}
\centering
\captionsetup[subfigure]{justification=centering}
\begin{subfigure}{.2\textwidth}
\centering
\resizebox{0.95\textwidth}{!}{%
\begin{tikzpicture}

\filldraw[fill=black!35!white] (0,1) rectangle (1,2);
\filldraw[fill=black!35!white] (1,3) rectangle (2,4);
\filldraw[fill=black!40!white] (2,1) rectangle (3,2);
\filldraw[fill=black!15!white] (3,2) rectangle (4,3);
\filldraw[fill=black!15!white] (4,1) rectangle (5,2);
\filldraw[fill=black!25!white] (1,0) rectangle (3,1);
\filldraw[fill=black!10!white] (3,3) rectangle (5,5);

\filldraw[fill=black!70!white] (0,3) rectangle (1,4);
\filldraw[fill=black!80!white] (1,2) rectangle (2,3);
\filldraw[fill=black!80!white] (2,4) rectangle (3,5);
\filldraw[fill=black!75!white] (3,3) rectangle (4,4);
\filldraw[fill=black!95!white] (3,1) rectangle (4,2);

\node[] at (-0.4,4.5) {$\mathbf{a}_1$};
\node[] at (-0.4,3.5) {$\mathbf{a}_2$};
\node[] at (-0.4,2.5) {$\mathbf{a}_3$};
\node[] at (-0.4,1.5) {$\mathbf{a}_4$};
\node[] at (-0.4,0.5) {$\mathbf{a}_5$};

\node[] at (0.5,5.4) {$\mathbf{b}_1$};
\node[] at (1.5,5.4) {$\mathbf{b}_2$};
\node[] at (2.5,5.4) {$\mathbf{b}_3$};
\node[] at (3.5,5.4) {$\mathbf{b}_4$};
\node[] at (4.5,5.4) {$\mathbf{b}_5$};

\draw[step=1cm,black,thick] (0,0) grid (5,5);
\end{tikzpicture}
}%
\caption{}
\label{cluster:weight}
\end{subfigure}\hspace{.05\linewidth}%
\begin{subfigure}{.2\textwidth}
\centering
\resizebox{0.95\textwidth}{!}{%
\begin{tikzpicture}
\filldraw[fill=black] (0,3) rectangle (1,4);
\filldraw[fill=black] (1,2) rectangle (2,3);
\filldraw[fill=black] (2,4) rectangle (3,5);
\filldraw[fill=black] (3,3) rectangle (4,4);
\filldraw[fill=black] (3,1) rectangle (4,2);

\node[] at (-0.4,4.5) {$\mathbf{a}_1$};
\node[] at (-0.4,3.5) {$\mathbf{a}_2$};
\node[] at (-0.4,2.5) {$\mathbf{a}_3$};
\node[] at (-0.4,1.5) {$\mathbf{a}_4$};
\node[] at (-0.4,0.5) {$\mathbf{a}_5$};

\node[] at (0.5,5.4) {$\mathbf{b}_1$};
\node[] at (1.5,5.4) {$\mathbf{b}_2$};
\node[] at (2.5,5.4) {$\mathbf{b}_3$};
\node[] at (3.5,5.4) {$\mathbf{b}_4$};
\node[] at (4.5,5.4) {$\mathbf{b}_5$};

\draw[step=1cm,black,thick] (0,0) grid (5,5);
\end{tikzpicture}
}%
\caption{}
\label{cluster:link}
\end{subfigure}\hspace{.05\linewidth}%
\begin{subfigure}{.2\textwidth}
\centering
\resizebox{0.95\textwidth}{!}{%
\begin{tikzpicture}
\filldraw[fill=blue!70!white] (0,3) rectangle (1,4);
\filldraw[fill=blue!70!white] (3,3) rectangle (4,4);
\filldraw[fill=blue!70!white] (3,1) rectangle (4,2);
\filldraw[fill=green!70!black] (1,2) rectangle (2,3);
\filldraw[fill=yellow!60!black] (2,4) rectangle (3,5);

\node[] at (0.5,3.5) {$\mathbf{1}$};
\node[] at (3.5,3.5) {$\mathbf{1}$};
\node[] at (3.5,1.5) {$\mathbf{1}$};
\node[] at (1.5,2.5) {$\mathbf{2}$};
\node[] at (2.5,4.5) {$\mathbf{3}$};

\node[] at (-0.4,4.5) {$\mathbf{a}_1$};
\node[] at (-0.4,3.5) {$\mathbf{a}_2$};
\node[] at (-0.4,2.5) {$\mathbf{a}_3$};
\node[] at (-0.4,1.5) {$\mathbf{a}_4$};
\node[] at (-0.4,0.5) {$\mathbf{a}_5$};

\node[] at (0.5,5.4) {$\mathbf{b}_1$};
\node[] at (1.5,5.4) {$\mathbf{b}_2$};
\node[] at (2.5,5.4) {$\mathbf{b}_3$};
\node[] at (3.5,5.4) {$\mathbf{b}_4$};
\node[] at (4.5,5.4) {$\mathbf{b}_5$};

\draw[step=1cm,black,thick] (0,0) grid (5,5);
\end{tikzpicture}
}%
\caption{}
\label{cluster:label}
\end{subfigure}\hspace{.05\linewidth}%
\begin{subfigure}{.2\textwidth}
\centering
\resizebox{0.95\textwidth}{!}{%
\begin{tikzpicture}
\filldraw[fill=blue!70!white] (1,3) rectangle (2,4);
\filldraw[fill=blue!70!white] (2,3) rectangle (3,4);
\filldraw[fill=blue!70!white] (2,2) rectangle (3,3);
\filldraw[fill=green!70!black] (3,1) rectangle (4,2);
\filldraw[fill=yellow!60!black] (4,0) rectangle (5,1);

\node[] at (2.5,3.5) {$\mathbf{1}$};
\node[] at (2.5,2.5) {$\mathbf{1}$};
\node[] at (1.5,3.5) {$\mathbf{1}$};
\node[] at (3.5,1.5) {$\mathbf{2}$};
\node[] at (4.5,0.5) {$\mathbf{3}$};

\node[] at (-0.4,4.5) {$\mathbf{a}_5$};
\node[] at (-0.4,3.5) {$\mathbf{a}_2$};
\node[] at (-0.4,2.5) {$\mathbf{a}_4$};
\node[] at (-0.4,1.5) {$\mathbf{a}_3$};
\node[] at (-0.4,0.5) {$\mathbf{a}_1$};

\node[] at (0.5,5.4) {$\mathbf{b}_5$};
\node[] at (1.5,5.4) {$\mathbf{b}_1$};
\node[] at (2.5,5.4) {$\mathbf{b}_4$};
\node[] at (3.5,5.4) {$\mathbf{b}_2$};
\node[] at (4.5,5.4) {$\mathbf{b}_3$};

\draw[step=1cm,black,thick] (0,0) grid (5,5);
\end{tikzpicture}
}%
\caption{}
\label{cluster:order}
\end{subfigure}
\caption{An example of \textit{post-hoc blocking} (\subref{cluster:weight}) shows an example of estimated weights with dark cells corresponding to larger weights. (\subref{cluster:link}) We convert all weights above a threshold $w_0$ to edges in adjacency matrix for a bipartite graph.  (\subref{cluster:label}) We number and color the connected components of the graph.  (\subref{cluster:order}) shows a reordering based on which connected component each node, corresponding to either a row or a column, is assigned to.  Nodes with no edges as assigned to a default cluster and left blank.}
\label{fig:cluster}
\end{figure*}
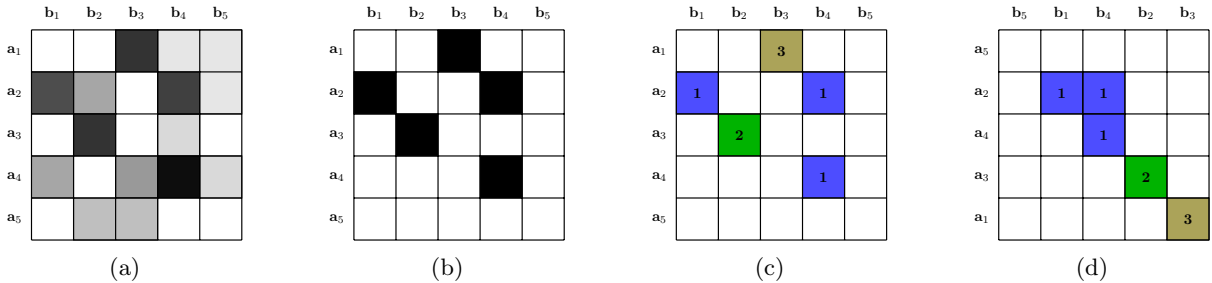

Reducing the number of record pairs under consideration is essential for scaling Bayesian PRL. Even traditional approaches to PRL almost always use a {\em blocking} scheme, for example only considering record pairs from the same geographic area as possible matches.  These geographic areas form {\em blocks} such that all the links between records occur withing the same block. Blocking is discussed in depth by \cite{herzog2007data, christen2012survey, christen2012data, steorts2014comparison}, among others.

We propose a new approach (that may be coupled with a first-pass of traditional blocking) which we call {\em post-hoc blocking}. The idea is to use the penalized likelihood estimates of the weights to filter out pairs that are unlikely to match (i.e., pairs with estimated weights below a threshold $w_0$), and simply ignore them during MCMC, defining a restricted version of the original MCMC algorithm. The procedure for obtaining post-hoc blocks is summarized in Algorithm~\ref{alg:posthoc}.

\begin{algorithm}
\caption{Post-hoc Blocking}
\label{alg:posthoc}
\begin{algorithmic}
\INPUT{Comparison vectors $\Gamma$, weight threshold $w_0$}
\OUTPUT{Penalized likeihood estimates of $m,u$, and $C$. An $n_A\times n_B$ adjacency matrix $G$, and its connected components defining post-hoc blocks. }
  \begin{enumerate}
  \item Generate $\hat m$, $\hat u$ and $\hat C$ via penalized maximum likelihood (Section~\ref{sec:penlike})
  \item Set $G = \ind{\log[\hat{m}(\gamma_{ab})] - \log[\hat{u}(\gamma_{ab})] >w_0}$
  \item Find the connected components of the bipartite graph with adjacency matrix $G$; these are the {\em post-hoc blocks}
  \end{enumerate}
%
%
%
\end{algorithmic}
\end{algorithm}

Figures~\ref{cluster:weight} - \ref{cluster:order} illustrate this process.  Figure ~\ref{cluster:weight} shows heatmap of the weights $\hat w_{ab} = \log[ \hat{m}(\gamma_{ab})/\hat{u}(\gamma_{ab})]$, with darker squares signifying larger weights. Figure~\ref{cluster:link} shows the thresholded matrix $G$, where the black entries are record pairs that could possibly be linked ($G_{ab}=1$) and the white entries (with $G_{ab}=0$) correspond to elements of $C$ that will be fixed at zero during MCMC. Figures~\ref{cluster:label} and~\ref{cluster:order} shows the connected components of $G$.  Finding the connected components of a bipartite graph is a well-studied problem with efficient solutions \citep{tarjan1972depth,gazit1986optimal}.  After thresholding, all the links must occur within these connected components (which we call {\em post-hoc blocks}).

This approach does not just reduce the number of record pairs under consideration --  it also makes the problem embarrassingly parallel.  Since we fix $C_{ab}=0$ anywhere $G_{ab}=0$, the entries of $C$ corresponding to each block (connected component of $G$) are conditionally independent of the remainder of $C$ given the other model parameters, and can be updated in parallel. Often $C$ is  updated using Metropolis-Hastings add/drop/swap link updates as described in e.g. \cite{green2006bayesian}.  If the blocks are small we have the option of jointly sampling all links within a block by enumerating all possibilities and doing a simple Gibbs update, providing further computational gains. 

\subsection{Choosing $w_0$}

Choosing the threshold $w_0$ requires balancing statistical accuracy against computational efficiency. Larger values of $w_0$ are more likely to exclude true matching pairs, increasing false non-match rates, and even excluding truly non-matching pairs which are not {\em obviously} non-matches risks misrepresenting posterior uncertainty.  Increasing $w_0$ tends to increase bias.

From a computational perspective, the number and size of the post-hoc blocks (and the total number of candidate record pairs to be visited during MCMC) are determined by the value of $w_0$.  A low threshold will admit a large set of record pairs split across fewer connected components in $G$. As $w_0$ increases, the number of candidate pairs decreases and the number of blocks increases at first as the elimination of ``weak ties'' leads existing blocks to split.  For large $w_0$ the number of blocks decreases as connected components of $G$ disappear instead of splitting into new components.

A natural strategy is to choose the smallest $w_0$ that leads to computationally feasible MCMC; this will naturally be context dependent. For example, in our large-scale example below, we chose the smallest $w_0$ with post-hoc blocks having at most 2500 record pairs.

\section{EXPERIMENTS}\label{sec:experiments}

To compare the performance of the three point estimates we introduce a simple Bayesian model with a posterior mode that coincides with the maximum of the penalized likelihood in \eqref{sec:penlike}.  This allows us to examine the differences between posterior modes and means when they are feasible to compute. However, our post-hoc blocking approach may be used in the context of any Bayesian model, and we do not advocate for this particular choice over others.

Our prior over $C$ is $p(C)\propto \exp(-\theta L)$ where $L = \sum_{a}\sum_{b} C_{ab}$.
\cite{green2006bayesian} originally introduced priors of this form for a related class of alignment problems. We take
\begin{align}
    (m_{j1}, \dots, m_{jk_j})&\sim Dir(\alpha_{mj1}, \dots,\alpha_{mjk_j})\\
    (u_{j1}, \dots, u_{jk_j})&\sim Dir(\alpha_{uj1}, \dots,\alpha_{ujk_j})
\end{align}
independently.

\vspace{-0.5em}
\subsection{Small-scale Example: Italian Census}\label{sec:italy}

We reanalyze data from the 2001 Italian census and a post-enumeration survey previously examined by \cite{tancredi2011hierarchical}. 
The data come from a small geographic area; there are 34 records from the census (file A) and 45 records from the post-enumeration survey (file B). The goal is to identify the number of overlapping records to obtain an estimate of the number of people missed by the census count using capture-recapture methods.

Each record includes three categorical variables: the first two consonants of the family name (339 categories), sex (2 categories), and education level (17 categories). We generate comparison vectors as binary indicators of an exact match between each field. We assume that $P(m_i)\sim \mathrm{Beta}(20, 3)$ for $i = 1, 2, 3$ and $P(u_j)\sim \mathrm{Beta}(3, 20)$ for $j = 1, 2, 3$.

\begin{figure*}
  \centering
  \begin{subfigure}{.223\linewidth}
    \centering
    \includegraphics[scale=0.45]{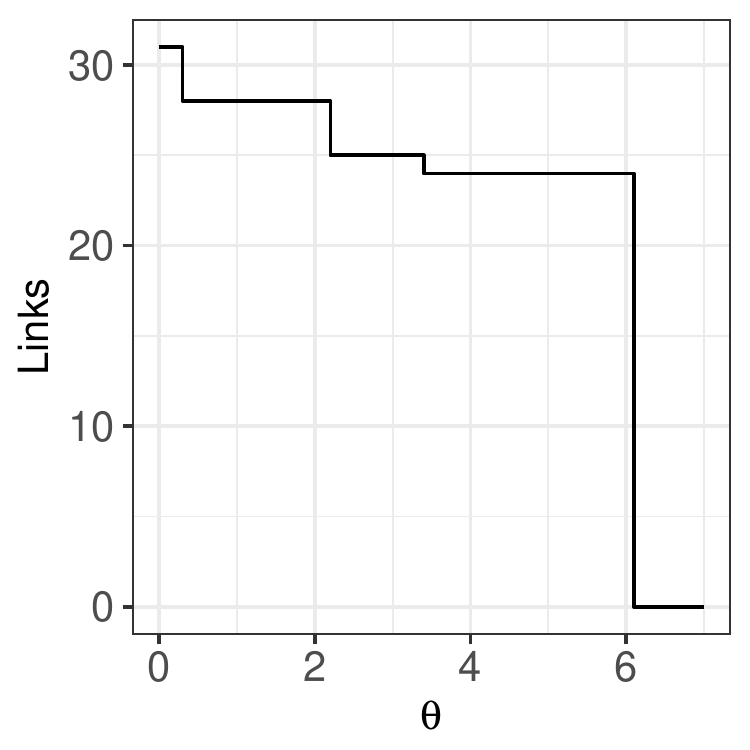}
    \caption{}
    \label{ic_results:MAP_link}
  \end{subfigure}\hspace{.01\linewidth}%
  \begin{subfigure}{.3\linewidth}
    \centering
    \includegraphics[scale=0.45]{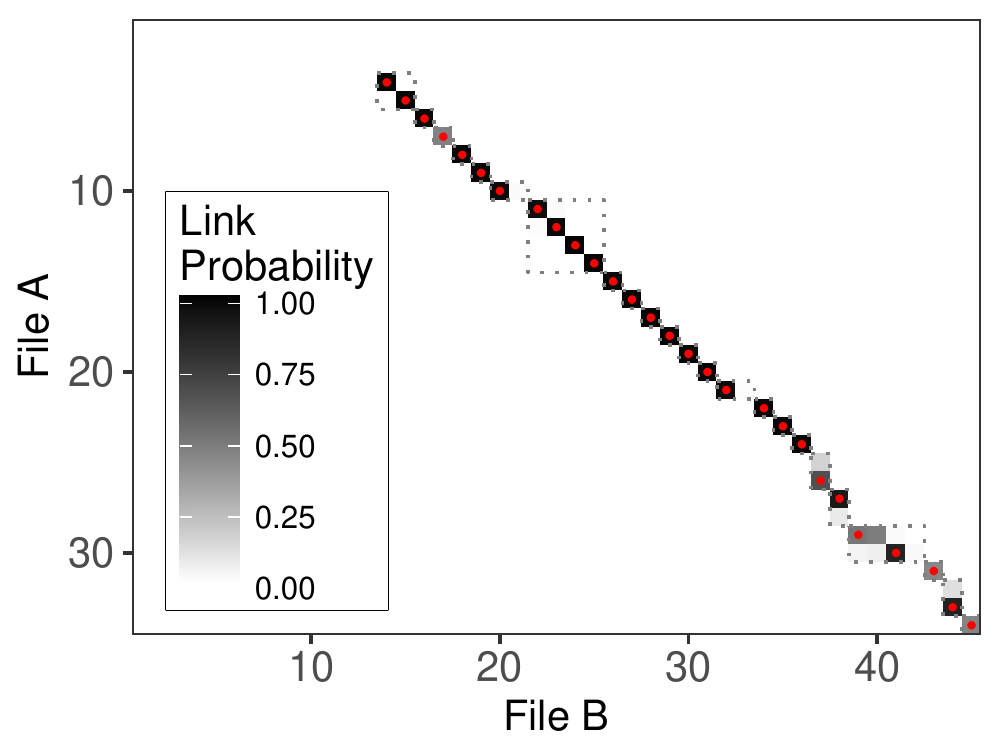}
    \caption{}
    \label{ic_results:MAP_dot}
  \end{subfigure}\hspace{.01\linewidth}%
  \begin{subfigure}{.223\linewidth}
    \centering
    \includegraphics[scale=0.45]{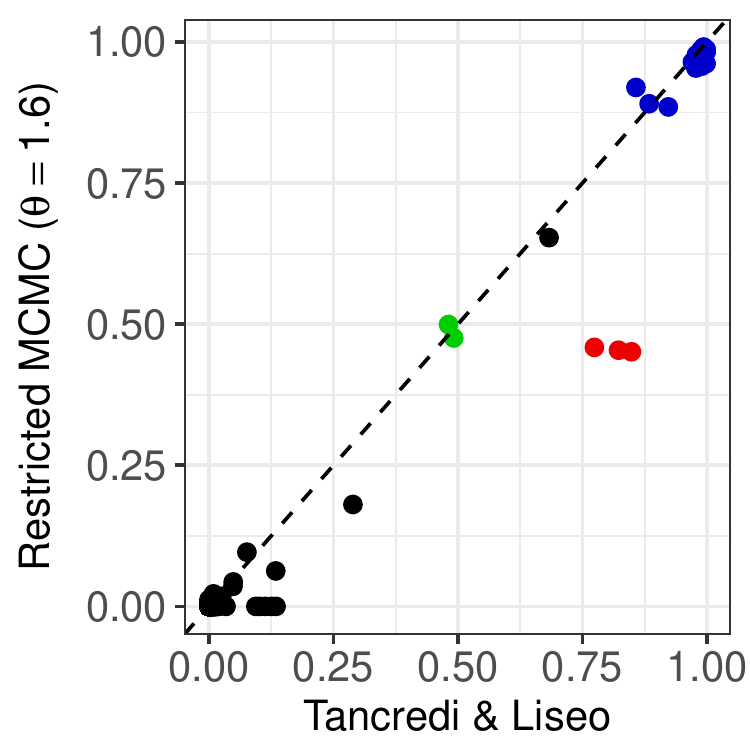}
    \caption{}
    \label{ic_results:tl_scatter}
  \end{subfigure}\hspace{.01\linewidth}%
  \begin{subfigure}{.223\linewidth}
    \centering
    \includegraphics[scale=0.45]{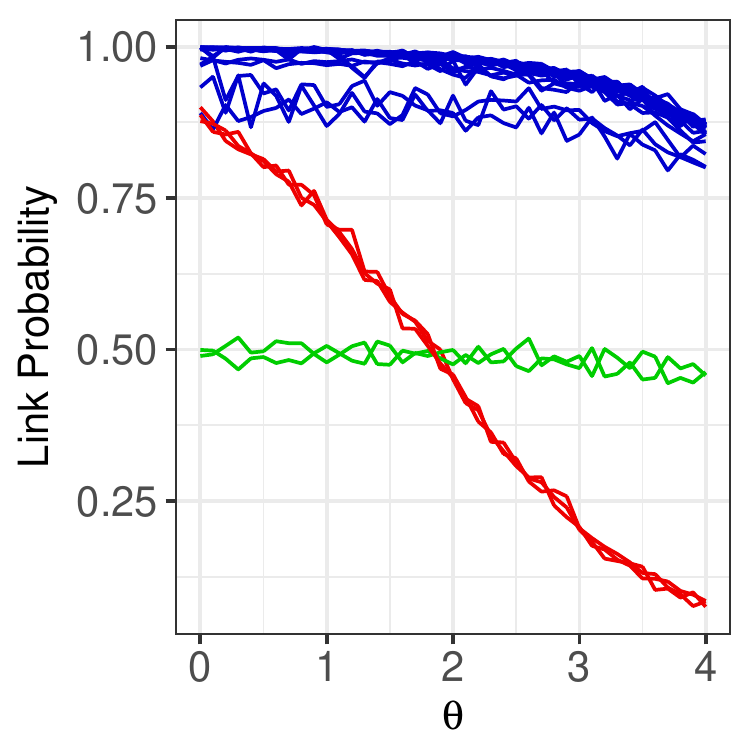}
    \caption{}
    \label{ic_results:sensitivity}
  \end{subfigure}
      \caption{(a) The number of links found by the penalized likelihood algorithm by choice of penalty parameter $\theta$. (b) Restricted MCMC posterior with penalized likelihood point estimate shown in red. (c) Posterior link probabilities from the restricted MCMC algorithm compared to those of Tancredi and Liseo. (d) The points for which the posterior link probability differs most between the approach of Tancredi and Liseo and the penalized likelihood approach are sensitive to the penalty value used in the prior.}
      \label{fig:ic_results}
\end{figure*}

\subsubsection{Penalized Likelihood Estimate}

We computed a series of penalized likelihood estimates across a range of $\theta$ values (0.0 to 7.0). The number of links estimated for each value of the penalty parameter is shown in Figure \ref{fig:ic_results}\subref{ic_results:MAP_link}; the estimate of $C$ is constant across the flat parts of the graph.  The entire runtime using the clue package \citep{clue2017manual} in R \citep{R} to solve the LSAP is about 0.3 seconds on a laptop with a 2.60 GHz processor.  

This procedure provides an efficient sensitivity analysis; using the results from Table 1 of \cite{tancredi2011hierarchical}, point estimates for the total population size range from 51 to 64, and the union of their 95\% credible intervals is $(49,78)$.  \cite{tancredi2011hierarchical} report a 95\% credible interval {\em under a single prior} of (49, 72), with a posterior median of 57. 

\begin{table}
  \centering
  \resizebox{\linewidth}{!}{%
  \begin{tabular}{rrrrrr}
    \hline
    \hline
    Last & Sex & Edu & Count & EM+LSAP Weight & Penalized Likelihood Weight\\
    \hline
    1 & 1 & 1 & 25 & 5.24 & 6.09 \\
    1 & 0 & 1 & 8 & 3.85 & 3.30 \\
    \rowcolor{Gray}
    1 & 1 & 0 & 13 & -7.49 & 2.10 \\
    \rowcolor{Gray}
    0 & 1 & 1 & 126 & 2.77 & -0.66 \\
    1 & 0 & 0 & 21 & -8.89 & -0.69 \\
    \rowcolor{Gray}
    0 & 0 & 1 & 78 & 1.37 & -3.44 \\
    0 & 1 & 0 & 601 & -9.96 & -4.64 \\
    0 & 0 & 0 & 658 & -11.36 & -7.43 \\
    \hline
    \hline
  \end{tabular}}
  \caption{Comparison between EM+LSAP and penalized likelihood weights for Italian census data with $\theta \in (0.3, 2.1)$.  Rows where the sign of the weight disagrees are shown in grey.}
  \label{tab:weights}
\end{table}

We also compared the penalized likelihood weight estimates (with $\theta = 2.0$) to the EM+LSAP values (Table~\ref{tab:weights}). There are several large discrepancies in gray. For example, the EM+LSAP weights imply that two records that agree on sex and education but not on the last name are {\em more} likely than not to be a match, while the penalized likelihood weight is negative. Counterintuitively, EM+LSAP also assigns a large negative weight to pairs agreeing on last name and sex but not education. While there is no ground truth for this dataset, these results make little sense -- as does the positive weight EM+LSAP assigns to record pairs that {\em only} agree on education.

\begin{figure}[t]
\centering
\includegraphics[scale=.35]{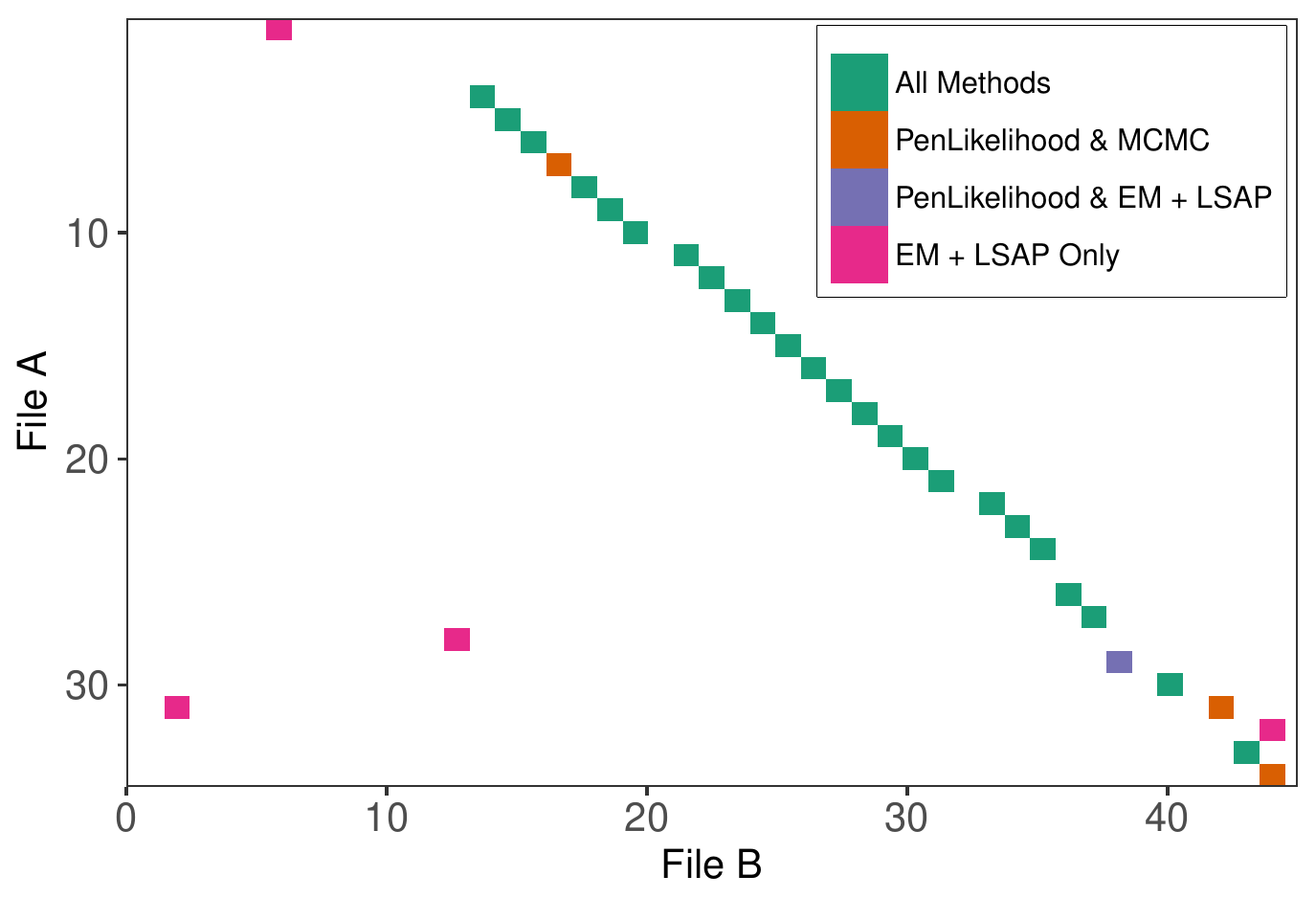}
\caption{Point estimates for links in the Italian census data. Green links are found by all methods, magenta are found only by the EM+LSAP approach, and so on.
The purple link is identified by the EM+LSAP and penalized likelihood approaches but not the MCMC algorithms because the row contains multiple identical comparison vectors, as discussed in Section \ref{sec:comp_point_est}.}
\label{fig:pointest}
\end{figure}

Figure~\ref{fig:pointest} shows that the penalized likelihood estimate closely follows the Bayes estimate derived from \cite{tancredi2011hierarchical}'s model as well as our own, providing further evidence of its superiority. The poor results are likely due to the EM+LSAP's weights obtained by maximizing the misspecified likelihood (which ignores one-to-one constraints). In contrast, the penalized likelihood accounts for one-to-one matching when estimating the weights, increasing accuracy.

\subsubsection{Restricted MCMC Estimation}
We ran the restricted MCMC algorithm using the penalized likelihood weights in Table~\ref{tab:weights} and a threshold of $w_0 = 0.0$, with $\theta = 2.0$ in the prior chosen based on the implied marginal prior over the number of links $L$. Figure \ref{fig:ic_results}\subref{ic_results:MAP_dot} shows the pairwise posterior probabilities of a match between each record pair, with the post-hoc blocks given by the dashed lines.  Only 46 (3\%) of the 1530 record pairs were above the threshold, significantly reducing the scale of the problem.

The posterior link probabilities are similar to those obtained using the significantly more complicated model introduced by \cite{tancredi2011hierarchical} (Figure \ref{fig:ic_results}\subref{ic_results:tl_scatter}).  The points shown in red are notable exceptions.  These three observations correspond to record pairs that match on the first two consonants of the family name and sex but not on level of education.  These links are marginal, in the sense that their posterior link probability is sensitive to the prior distribution over the total number of links. Figure \ref{fig:ic_results}\subref{ic_results:sensitivity} shows that as we increase $\theta$ in the prior, the marginal probability of linking these records drops precipitously.  In contrast, the posterior probability of linking record pairs shown in blue and green show little sensitivity to the choice of prior.  

\begin{figure*}[t]
  \centering
  \begin{subfigure}{.5\linewidth}
    \includegraphics[scale=.6]{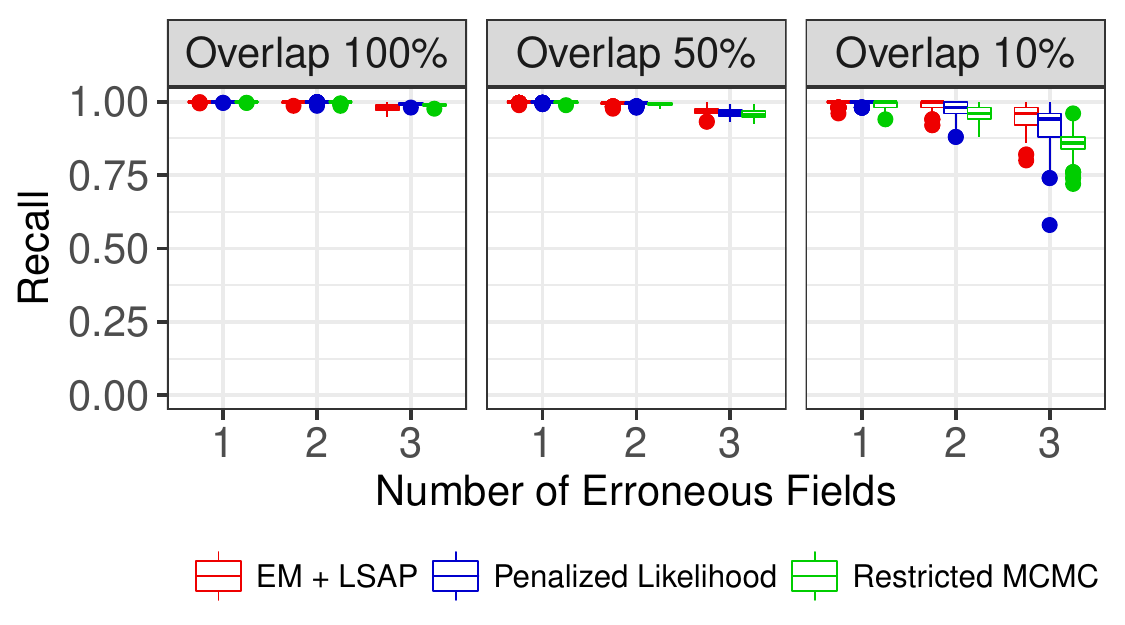}
  \end{subfigure}%
  \begin{subfigure}{.5\linewidth}
    \includegraphics[scale=.6]{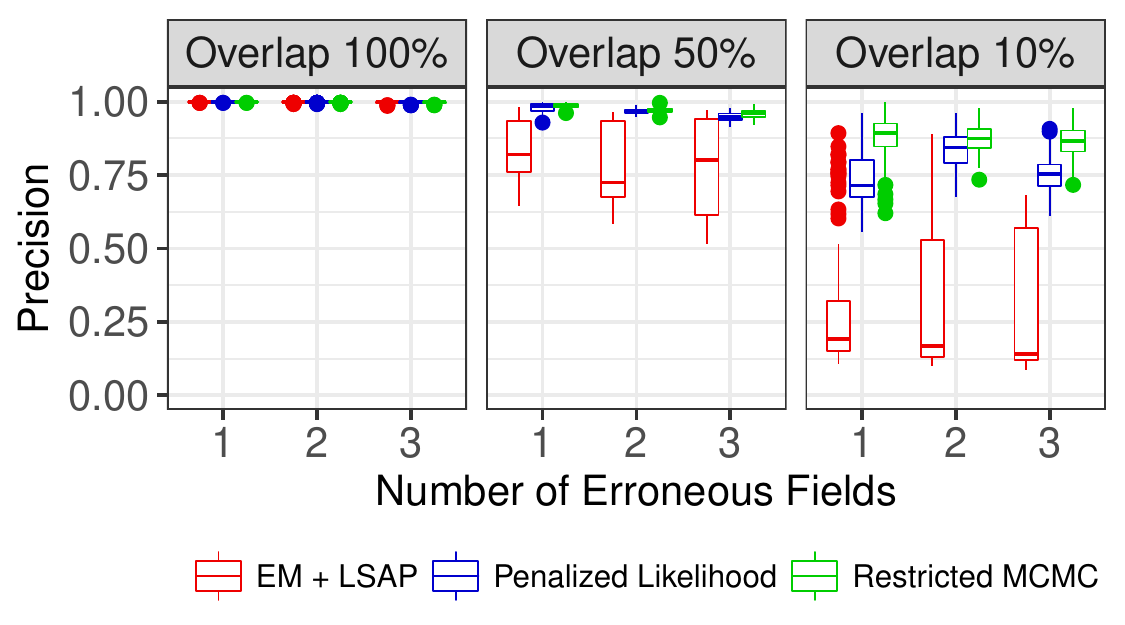}
  \end{subfigure}
  \caption{Recall 
  (left) and precision (right) 
  of the three point estimates in Section~\ref{sec:simstudy}. 
  }
  \label{fig:simperformance}
\end{figure*}

\subsection{Simulation Study}\label{sec:simstudy}

We use synthetic data provided by \cite{sadinle2016bayesian} for a simulation study.  Each synthetic dataset comprises two files of 500 records ($n_A = 500$, $n_B = 500$) taken from one of 100 large databases.  Each record contains four fields: given name, family name, age and occupation categories. We run simulations for each data file where errors are introduced into 1, 2 or 3 of the 4 fields for each record and the share of records which are linked is 100\%, 50\% or 10\%, for a total of 900 simulations.   Given name and family name are compared based on a discretized Levenshtein distance measure with four levels: exact agreement (distance of zero); mild disagreement, (0, .25]; moderate disagreement, (.25, .50]; and extreme disagreement, (.50, 1.0].  Age and occupation categories were compared using exact agreement.  See \cite{sadinle2016bayesian} for further details.  

For the penalized likelihood estimator and restricted MCMC, we vary $\theta$ by the percent of records which correspond to a true match.  We use values of 0.0, 5.0, and 7.0 for overlap percentages of 100\%, 50\% and 10\% respectively.  For 10\% and 50\% overlap these roughly correspond to prior distributions with a mode at a higher number of links than exist in the data, to mimic a conservative approach. (In practice, with a single dataset we would recommend a sensitivity analysis).  We take $m_{j}\sim Dir(1, 2, 5, 10)$ and $p(u_j)\propto 1$ for the discretized string comparisons.

We took $w_0=0$ in the restricted MCMC.  The restricted MCMC algorithm is run for 1000 steps with the first 100 discarded as burn-in, the same length of MCMC chain as \cite{sadinle2016bayesian}, but operating over a smaller state space.  The Bayes estimate is computed according to \eqref{eq:bayesest}.

Figure \ref{fig:simperformance} shows precision and recall for each of the three estimators in each scenario.  Performance degrades with more errors per record, but far less for the penalized likelihood point estimate and the Bayes estimate using restricted MCMC than for EM+LSAP. Inspecting the results we find that EM+LSAP frequently suffers from poorly estimated weights like in the Italian census example. This occurs in up to half of the datasets with 10\% overlap. 

Note that unlike in the Italian census example, here we have a measure of ground truth for the parameters (in terms of the empirical conditional distributions of $\gamma$ given true match and non-match status).  Using this information to initialize the EM algorithm fails to remedy the problem (as does repeated random initialization), suggesting that these are not just poor local modes. The EM estimates are too unreliable for estimating C via EM+LSAP or for post-hoc blocking.

The performance of the post-hoc blocked MCMC algorithm using the penalized likelihood weights is comparable to the results presented by \cite{sadinle2016bayesian} using an alternative prior for $C$.  Comparing the penalized likelihood estimates to the Bayes estimates, in the 10\% overlap scenario the restricted MCMC algorithm generally has higher precision with slightly lower recall. This is largely a function of the Bayes decision rule, where a record from file $A$ with multiple plausible candidates in file $B$ has its posterior probability ``smeared'' across all the candidates, compared to the behavior of the penalized likelihood estimator, which will choose the best possible link with weight over the threshold (breaking any ties randomly). 

\subsection{Large-scale Synthetic Example}

We constructed a large-scale synthetic dataset by taking all 100 data files from Section \ref{sec:simstudy} and stacking them to two datasets with $50,000$ records each, for 2.5 billion record pairs.  We examine the two erroneous field/50\% overlap (25,000 truly matching pair) case. 
It is infeasible to compute the comparisons vectors for 2.5 billion record pairs, so as an initial traditional blocking step we exclude any record pairs that have different postal codes.  The 29 resulting blocks each contain between two and 33 million records pairs.  About 160 million of the 2.5 billion record pairs remain after blocking.  We ran the penalized likelihood and the EM+LSAP algorithms with $\theta=7.5$, solving the LSAP problem in parallel across the blocks.  

The EM+LSAP estimate of $C$ has a recall of 97.1\% but a precision of only 62.6\%; it is prone to making erroneous links whatever the cutoff used. In contrast, the penalized likelihood estimate with $\theta=7.5$ has a recall of 92.7\% and precision of 94.6\%.  The EM+LSAP approach has a total run time of 320 seconds with the EM algorithm taking 116 seconds to converge and the LSAP taking a further 304 seconds to solve.  Despite repeated calls to the LSAP solver the penalized likelihood approach is faster, taking a total of 290 seconds.  

Finally, we used the penalized likelihood weights to perform post-hoc blocking with $w_0=4.9$.  This threshold was selected to limit the size of the post-hoc blocks to no more than 50 records from either file, yielding 23,903 total blocks. The number of pairs under consideration reduces to 36,356 from over 160 million, for a reduction ratio (proportion of record pairs excluded by post-hoc blocking) of about 99.98\%. Of the 25,000 truly matching record pairs 24,877 are included in the same post-hoc block, for a pairs completeness (proportion of truly matching record pairs retained by post-hoc blocking) of $24877/25000\approx$99.5\%.  The majority of the post-hoc blocks (19,901) contain only a single record pair while the largest contains 32 and 31 records from file A and B, respectively.  

\begin{figure}[t]
  \centering
  \begin{subfigure}{.5\linewidth}
    \includegraphics[scale=.35]{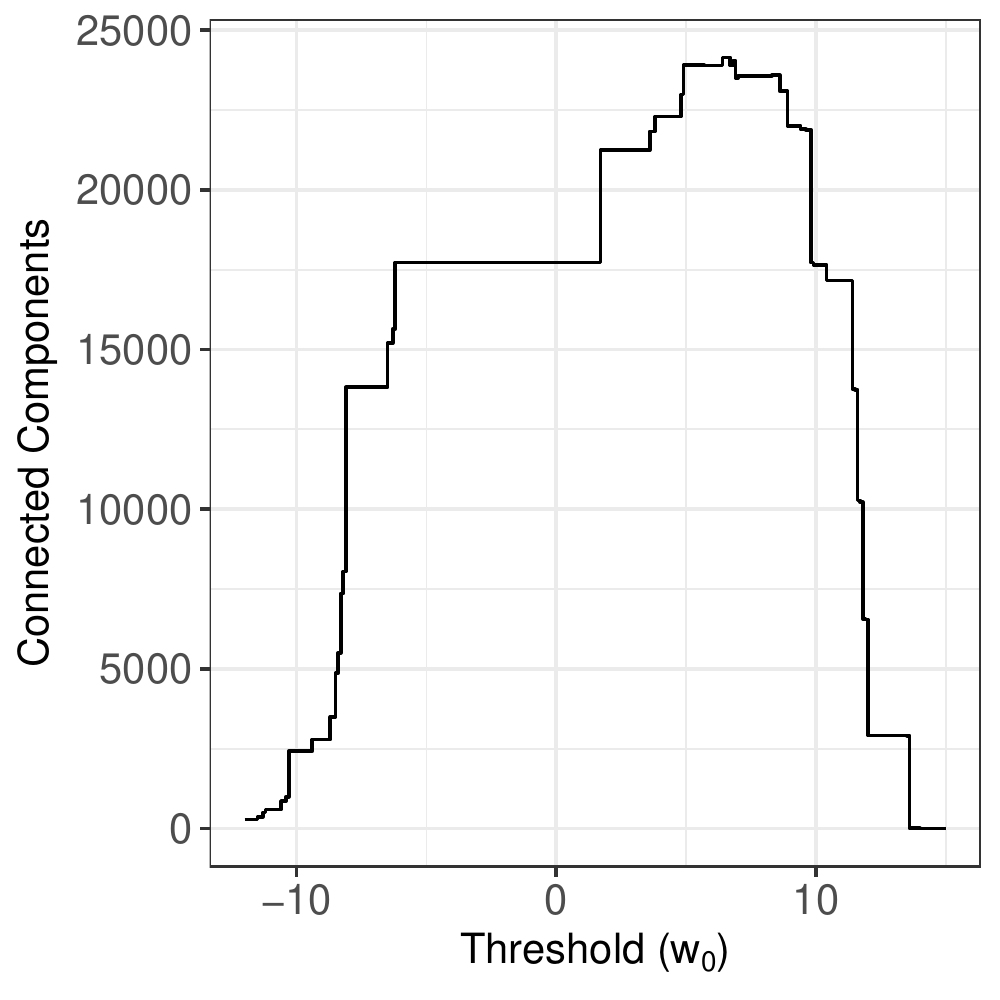}
    \caption{}
    \label{components:ncomponents}
  \end{subfigure}%
  \begin{subfigure}{.5\linewidth}
    \includegraphics[scale=.35]{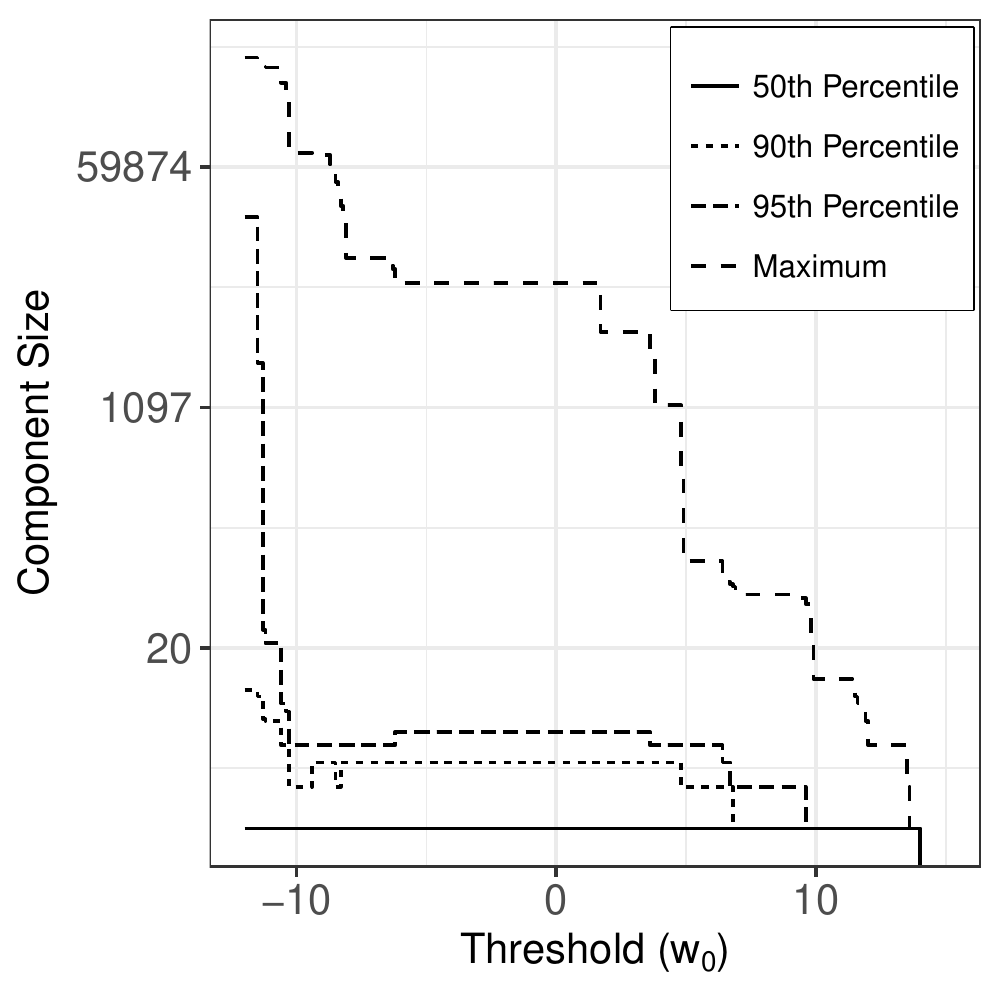}
    \caption{}
    \label{components:size}
  \end{subfigure}
  \begin{subfigure}{.5\linewidth}
    \includegraphics[scale=.35]{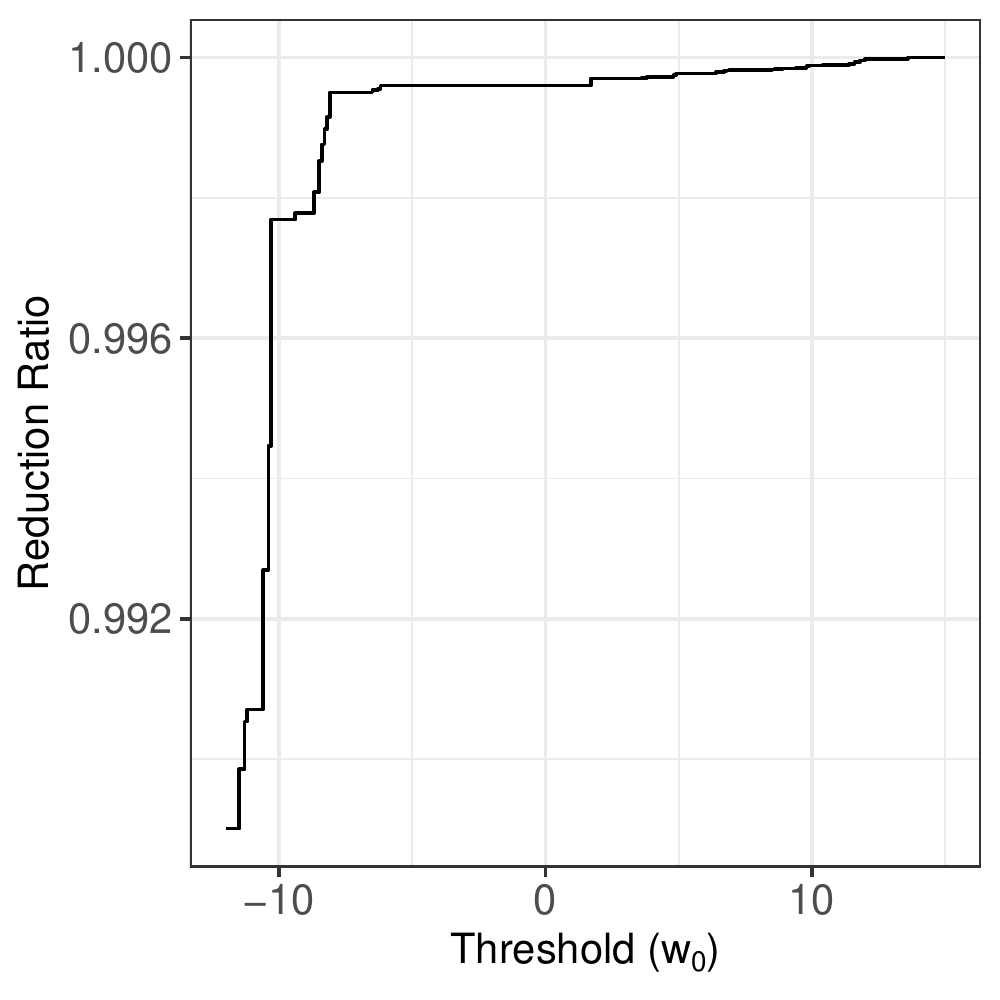}
    \caption{}
    \label{components:rr}
  \end{subfigure}%
    \begin{subfigure}{.5\linewidth}
    \includegraphics[scale=.35]{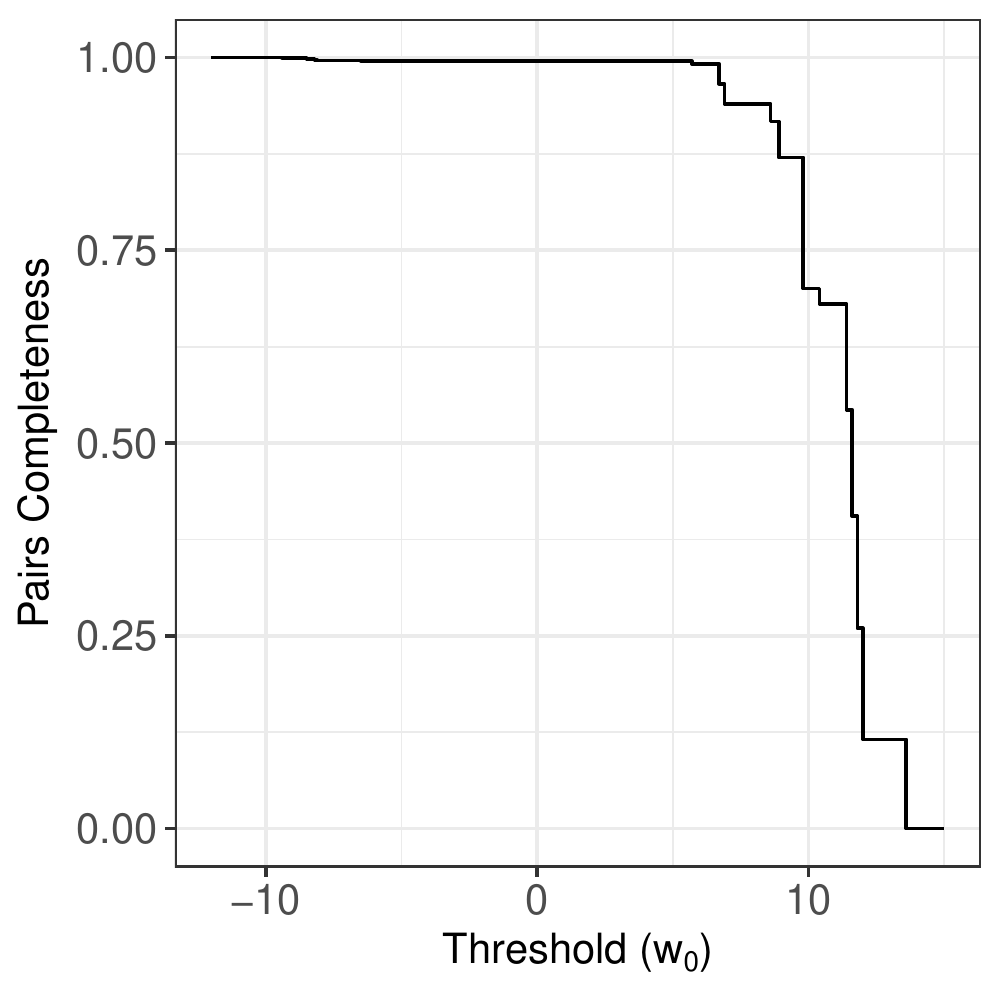}
    \caption{}
    \label{components:pc}
  \end{subfigure}
  \caption{Post-hoc blocks resulting from varying threshold value $w_0$. (\subref{components:ncomponents}) Number of blocks. (\subref{components:size}) Block size quantiles. (\subref{components:rr}) Reduction ratio (proportion of record pairs excluded by post-hoc blocking). (\subref{components:pc}) Pairs completeness (proportion of truly matching record pairs retained by post-hoc blocking).}
  \label{fig:components}
\end{figure}

The effect of $w_0$ on the size of the post-hoc blocks, the reduction ratio, and the pairs completeness metric is shown in Figure \ref{fig:components}. There are a range of values for $w_0$ (roughly $\pm 5$) that appear to make reasonable tradeoffs between computational gains and false non-matches. 

We run our restricted MCMC algorithm for 2,500 iterations (where each iteration performs an add/delete/swap move within each of the $23,903$ post-hoc blocks, which may be trivially parallelized) after 1,000 iterations of burn-in.   The Bayes estimate improves slightly on the recall rate of the penalized likelihood estimate (95.5\%) with precision of 94.6\%.  More importantly, the restricted MCMC provides approximate posterior samples, and the potential for much richer inferences and propagation of uncertainty through to subsequent analysis using the linked files. 

\section{CONCLUSION}
We present new approaches to performing Bayesian record linkage with considerable advantages over previous approaches.  We provide a computationally efficient method for point estimation by jointly estimating model parameters and linkage structure via penalized likelihood. Using this point estimate we introduce {\em post-hoc blocking}, a simple and efficient method for dramatically enhancing the computational efficiency of MCMC algorithms by targeting only the links that have significant posterior uncertainty.

There is significant existing work on specifying models for Bayesian record linkage, including those that model record features directly rather than via comparisons  (see e.g. \cite{gutman2013bayesian, fortini2002modelling, tancredi2011hierarchical, steorts2015entity, steorts2016bayesian}).  Our post-hoc blocking and restricted MCMC is readily applied there as well (penalized likelihood weights need not correspond to the posterior mode of the Bayesian model under consideration). 

Generalizing this approach to three or more files and duplicate records is challenging; the models involved become more complicated, the natural analogue of the penalized likelihood is less obvious, and the optimization problems become more difficult to solve. We expect this to be a fruitful line of research as Bayesian methods for probabilistic record linkage are deployed in more applied problems involving larger datasets.

\newpage
\bibliographystyle{apa}
\bibliography{rl_citations.bib}

\begin{thebibliography}{}

\bibitem[\protect\astroncite{Alicandro et~al.}{2017}]{alicandro2017differences}
Alicandro, G., Frova, L., Sebastiani, G., Boffetta, P., and La~Vecchia, C.
  (2017).
\newblock Differences in education and premature mortality: a record linkage
  study of over 35 million italians.
\newblock {\em European Journal of Public Health}.

\bibitem[\protect\astroncite{Christen}{2012a}]{christen2012data}
Christen, P. (2012a).
\newblock {\em Data matching: concepts and techniques for record linkage,
  entity resolution, and duplicate detection}.
\newblock Springer Science \& Business Media.

\bibitem[\protect\astroncite{Christen}{2012b}]{christen2012survey}
Christen, P. (2012b).
\newblock A survey of indexing techniques for scalable record linkage and
  deduplication.
\newblock {\em IEEE transactions on knowledge and data engineering},
  24(9):1537--1555.

\bibitem[\protect\astroncite{Dalzell and Reiter}{2016}]{dalzell2016regression}
Dalzell, N.~M. and Reiter, J.~P. (2016).
\newblock Regression modeling and file matching using possibly erroneous
  matching variables.
\newblock {\em arXiv preprint arXiv:1608.06309}.

\bibitem[\protect\astroncite{Dusetzina et~al.}{2014}]{dusetzina2014linking}
Dusetzina, S.~B., Tyree, S., Meyer, A.-M., Meyer, A., Green, L., and Carpenter,
  W.~R. (2014).
\newblock {\em Linking data for health services research: a framework and
  instructional guide}.
\newblock Agency for Healthcare Research and Quality (US), Rockville (MD).

\bibitem[\protect\astroncite{Enamorado et~al.}{2017}]{enamorado2017using}
Enamorado, T., Fifield, B., and Imai, K. (2017).
\newblock Using a probabilistic model to assist merging of large-scale
  administrative records.

\bibitem[\protect\astroncite{Fellegi and Sunter}{1969}]{fellegi1969theory}
Fellegi, I.~P. and Sunter, A.~B. (1969).
\newblock A theory for record linkage.
\newblock {\em Journal of the American Statistical Association},
  64(328):1183--1210.

\bibitem[\protect\astroncite{Fortini et~al.}{2002}]{fortini2002modelling}
Fortini, M., Nuccitelli, A., Liseo, B., and Scanu, M. (2002).
\newblock Modelling issues in record linkage: a bayesian perspective.
\newblock In {\em Proceedings of the American Statistical Association, Survey
  Research Methods Section}, pages 1008--1013.

\bibitem[\protect\astroncite{Gazit}{1986}]{gazit1986optimal}
Gazit, H. (1986).
\newblock An optimal randomized parallel algorithm for finding connected
  components in a graph.
\newblock In {\em Foundations of Computer Science, 1986., 27th Annual Symposium
  on}, pages 492--501. IEEE.

\bibitem[\protect\astroncite{Green}{2015}]{green2015mad}
Green, P.~J. (2015).
\newblock Mad-bayes matching and alignment for labelled and unlabelled
  configurations.
\newblock {\em Geometry Driven Statistics}, 121:377.

\bibitem[\protect\astroncite{Green and Mardia}{2006}]{green2006bayesian}
Green, P.~J. and Mardia, K.~V. (2006).
\newblock Bayesian alignment using hierarchical models, with applications in
  protein bioinformatics.
\newblock {\em Biometrika}, 93(2):235--254.

\bibitem[\protect\astroncite{Gu and Gutman}{2016}]{gu2016combining}
Gu, C. and Gutman, R. (2016).
\newblock Combining item response theory with multiple imputation to equate
  health assessment questionnaires.
\newblock {\em Biometrics}.

\bibitem[\protect\astroncite{Gutman et~al.}{2013}]{gutman2013bayesian}
Gutman, R., Afendulis, C.~C., and Zaslavsky, A.~M. (2013).
\newblock A bayesian procedure for file linking to analyze end-of-life medical
  costs.
\newblock {\em Journal of the American Statistical Association},
  108(501):34--47.

\bibitem[\protect\astroncite{Herzog et~al.}{2007}]{herzog2007data}
Herzog, T.~N., Scheuren, F.~J., and Winkler, W.~E. (2007).
\newblock {\em Data quality and record linkage techniques}.
\newblock Springer Science \& Business Media.

\bibitem[\protect\astroncite{Hornik}{2017}]{clue2017manual}
Hornik, K. (2017).
\newblock {\em clue: Cluster ensembles}.
\newblock R package version 0.3-54.

\bibitem[\protect\astroncite{Jaro}{1989}]{jaro1989advances}
Jaro, M.~A. (1989).
\newblock Advances in record-linkage methodology as applied to matching the
  1985 census of tampa, florida.
\newblock {\em Journal of the American Statistical Association},
  84(406):414--420.

\bibitem[\protect\astroncite{Jonker and Volgenant}{1986}]{jonker1986improving}
Jonker, R. and Volgenant, T. (1986).
\newblock Improving the hungarian assignment algorithm.
\newblock {\em Operations Research Letters}, 5(4):171--175.

\bibitem[\protect\astroncite{Kuhn}{1955}]{kuhn1955hungarian}
Kuhn, H.~W. (1955).
\newblock The hungarian method for the assignment problem.
\newblock {\em Naval Research Logistics (NRL)}, 2(1-2):83--97.

\bibitem[\protect\astroncite{Lawler}{1976}]{lawler1976combinatorial}
Lawler, E.~L. (1976).
\newblock {\em Combinatorial optimization: networks and matroids}.
\newblock Courier Corporation.

\bibitem[\protect\astroncite{Liseo and Tancredi}{2011}]{liseo2011bayesian}
Liseo, B. and Tancredi, A. (2011).
\newblock Bayesian estimation of population size via linkage of multivariate
  normal data sets.
\newblock {\em Journal of Official Statistics}, 27(3):491--505.

\bibitem[\protect\astroncite{Mackay et~al.}{2015}]{mackay2015educational}
Mackay, D.~F., Wood, R., King, A., Clark, D.~N., Cooper, S.-A., Smith, G.~C.,
  and Pell, J.~P. (2015).
\newblock Educational outcomes following breech delivery: a record-linkage
  study of 456 947 children.
\newblock {\em International journal of epidemiology}, 44(1):209--217.

\bibitem[\protect\astroncite{Neter et~al.}{1965}]{neter1965effect}
Neter, J., Maynes, E.~S., and Ramanathan, R. (1965).
\newblock The effect of mismatching on the measurement of response errors.
\newblock {\em Journal of the American Statistical Association},
  60(312):1005--1027.

\bibitem[\protect\astroncite{{R Core Team}}{2017}]{R}
{R Core Team} (2017).
\newblock {\em R: A Language and Environment for Statistical Computing}.
\newblock R Foundation for Statistical Computing, Vienna, Austria.

\bibitem[\protect\astroncite{Sadinle}{2017}]{sadinle2016bayesian}
Sadinle, M. (2017).
\newblock Bayesian estimation of bipartite matchings for record linkage.
\newblock {\em Journal of the American Statistical Association},
  112(518):600--612.

\bibitem[\protect\astroncite{Sadinle et~al.}{2014}]{sadinle2014detecting}
Sadinle, M. et~al. (2014).
\newblock Detecting duplicates in a homicide registry using a bayesian
  partitioning approach.
\newblock {\em The Annals of Applied Statistics}, 8(4):2404--2434.

\bibitem[\protect\astroncite{Sauleau et~al.}{2005}]{sauleau2005medical}
Sauleau, E.~A., Paumier, J.-P., and Buemi, A. (2005).
\newblock Medical record linkage in health information systems by approximate
  string matching and clustering.
\newblock {\em BMC Medical Informatics and Decision Making}, 5(1):32.

\bibitem[\protect\astroncite{Steorts et~al.}{2015}]{steorts2015entity}
Steorts, R.~C. et~al. (2015).
\newblock Entity resolution with empirically motivated priors.
\newblock {\em Bayesian Analysis}, 10(4):849--875.

\bibitem[\protect\astroncite{Steorts et~al.}{2016}]{steorts2016bayesian}
Steorts, R.~C., Hall, R., and Fienberg, S.~E. (2016).
\newblock A bayesian approach to graphical record linkage and deduplication.
\newblock {\em Journal of the American Statistical Association},
  111(516):1660--1672.

\bibitem[\protect\astroncite{Steorts et~al.}{2014}]{steorts2014comparison}
Steorts, R.~C., Ventura, S.~L., Sadinle, M., and Fienberg, S.~E. (2014).
\newblock A comparison of blocking methods for record linkage.
\newblock In {\em International Conference on Privacy in Statistical
  Databases}, pages 253--268. Springer.

\bibitem[\protect\astroncite{Tancredi et~al.}{2013}]{tancredi2013accounting}
Tancredi, A., Auger-M{\'e}th{\'e}, M., Marcoux, M., and Liseo, B. (2013).
\newblock Accounting for matching uncertainty in two stage capture--recapture
  experiments using photographic measurements of natural marks.
\newblock {\em Environmental and ecological statistics}, 20(4):647--665.

\bibitem[\protect\astroncite{Tancredi et~al.}{2011}]{tancredi2011hierarchical}
Tancredi, A., Liseo, B., et~al. (2011).
\newblock A hierarchical bayesian approach to record linkage and population
  size problems.
\newblock {\em The Annals of Applied Statistics}, 5(2B):1553--1585.

\bibitem[\protect\astroncite{Tarjan}{1972}]{tarjan1972depth}
Tarjan, R. (1972).
\newblock Depth-first search and linear graph algorithms.
\newblock {\em SIAM journal on computing}, 1(2):146--160.

\bibitem[\protect\astroncite{Ventura and
  Nugent}{2014}]{ventura2014hierarchical}
Ventura, S.~L. and Nugent, R. (2014).
\newblock Hierarchical linkage clustering with distributions of distances for
  large-scale record linkage.
\newblock In {\em International Conference on Privacy in Statistical
  Databases}, pages 283--298. Springer.

\bibitem[\protect\astroncite{Winkler}{1988}]{winkler1988using}
Winkler, W.~E. (1988).
\newblock Using the em algorithm for weight computation in the fellegi-sunter
  model of record linkage.
\newblock In {\em Proceedings of the Section on Survey Research Methods,
  American Statistical Association}, volume 667, page 671.

\bibitem[\protect\astroncite{Winkler and
  Thibaudeau}{1991}]{winkler1991application}
Winkler, W.~E. and Thibaudeau, Y. (1991).
\newblock An application of the fellegi-sunter model of record linkage to the
  1990 us decennial census.
\newblock {\em US Bureau of the Census}, pages 1--22.

\end{thebibliography}

\newpage \quad
\newpage
\appendix
\section{LSAP}\label{sec:LSAP}
The optimization problem in \eqref{eq:penlike} of Section~\ref{sec:penlike} is solved using a LSAP based on modified weights \begin{equation} \label{eq:pweight}
\tilde w_{ab} = \left\{\begin{array}{lr}
w_{ab} - \theta & w_{ab} - \theta \geq 0\\
0 & w_{ab} < \theta
\end{array}\right.,
\end{equation}
the result of applying soft-thresholding to the weights defined in \eqref{eq:weight}.  We then solve the optimization problem:
\begin{equation} \label{eq:truncLSAP}
\begin{aligned}
\max_{C}   &\sum_{a,b\in A\times B} C_{ab} \tilde w_{ab} \\
\subjectto\quad  & C_{ab} \in\{0, 1\}\\
&\sum_{b \in B} \hspace{0.5em} C_{ab} \leq 1 \hspace{2em} \forall a\in A \\
&\sum_{a \in A} \hspace{0.5em} C_{ab} \leq 1 \hspace{2em} \forall b\in B.
\end{aligned}
\end{equation}
If $C^*$ is the value of $C$ which maximizes \eqref{eq:penlike} then $C^*$ also maximizes \eqref{eq:truncLSAP} as the coefficients of all entries of $C_{ab}$ for which $\tilde w_{ab} > 0$ are unaffected by the transformation.  However, the same objective value will be achieved by any value of $C$ where $C_{ab} = 1$ for all $a,b$ where $C_{ab}^* = 1$ but there may also exist $a,b$ where $C_{ab} = 1$ and $C_{ab}^*$ = 0.  It is this equivalence which allow the optimal solution to be found using by solving a LSAP, which assigns all rows to a column.   The soft-thresholding ensures that optimal objective value of the LSAP will be identical to that of \eqref{eq:penlike} allowing $C^*$ to be recovered by simply deleting links where $\tilde w_{ab} = 0$.

\end{document}